\documentclass[sigplan,10pt]{acmart}
\pdfoutput=1

\renewcommand\footnotetextcopyrightpermission[1]{}
\usepackage{listings}     
\usepackage{booktabs}     
\usepackage{array}        
\usepackage{tabularx}     
\usepackage{enumitem}     

\usepackage{comment}

\usepackage[utf8]{inputenc}
\usepackage[T1]{fontenc}
\usepackage{textcomp}
\usepackage[english]{babel} 

\DeclareUnicodeCharacter{00A0}{~}      
\DeclareUnicodeCharacter{00A1}{\textexclamdown}
\DeclareUnicodeCharacter{00A2}{\textcent}
\DeclareUnicodeCharacter{00A3}{\textsterling}
\DeclareUnicodeCharacter{00A5}{\textyen}
\DeclareUnicodeCharacter{00A7}{\S}
\DeclareUnicodeCharacter{00A8}{\"{}}
\DeclareUnicodeCharacter{00A9}{\textcopyright}
\DeclareUnicodeCharacter{00AA}{\textordfeminine}
\DeclareUnicodeCharacter{00AB}{\guillemotleft}
\DeclareUnicodeCharacter{00AC}{\textlnot}
\DeclareUnicodeCharacter{00AD}{}       
\DeclareUnicodeCharacter{00AE}{\textregistered}
\DeclareUnicodeCharacter{00AF}{\={}}   

\DeclareUnicodeCharacter{00B0}{\textdegree}
\DeclareUnicodeCharacter{00B1}{\textpm}
\DeclareUnicodeCharacter{00B2}{\textsuperscript{2}}
\DeclareUnicodeCharacter{00B3}{\textsuperscript{3}}
\DeclareUnicodeCharacter{00B4}{\'{}}   
\DeclareUnicodeCharacter{00B5}{\textmu}
\DeclareUnicodeCharacter{00B6}{\P}
\DeclareUnicodeCharacter{00B7}{\textperiodcentered}
\DeclareUnicodeCharacter{00B8}{\c{}}   
\DeclareUnicodeCharacter{00B9}{\textsuperscript{1}}
\DeclareUnicodeCharacter{00BA}{\textordmasculine}
\DeclareUnicodeCharacter{00BB}{\guillemotright}
\DeclareUnicodeCharacter{00BC}{\textonequarter}
\DeclareUnicodeCharacter{00BD}{\textonehalf}
\DeclareUnicodeCharacter{00BE}{\textthreequarters}
\DeclareUnicodeCharacter{00BF}{\textquestiondown}

\DeclareUnicodeCharacter{00C0}{\`A}
\DeclareUnicodeCharacter{00C1}{\'A}
\DeclareUnicodeCharacter{00C2}{\^A}
\DeclareUnicodeCharacter{00C3}{\~A}
\DeclareUnicodeCharacter{00C4}{\"A}
\DeclareUnicodeCharacter{00C5}{\AA}
\DeclareUnicodeCharacter{00C6}{\AE}
\DeclareUnicodeCharacter{00C7}{\c{C}}
\DeclareUnicodeCharacter{00C8}{\`E}
\DeclareUnicodeCharacter{00C9}{\'E}
\DeclareUnicodeCharacter{00CA}{\^E}
\DeclareUnicodeCharacter{00CB}{\"E}
\DeclareUnicodeCharacter{00CC}{\`I}
\DeclareUnicodeCharacter{00CD}{\'I}
\DeclareUnicodeCharacter{00CE}{\^I}
\DeclareUnicodeCharacter{00CF}{\"I}
\DeclareUnicodeCharacter{00D1}{\~N}
\DeclareUnicodeCharacter{00D2}{\`O}
\DeclareUnicodeCharacter{00D3}{\'O}
\DeclareUnicodeCharacter{00D4}{\^O}
\DeclareUnicodeCharacter{00D5}{\~O}
\DeclareUnicodeCharacter{00D6}{\"O}
\DeclareUnicodeCharacter{00D8}{\O}
\DeclareUnicodeCharacter{00D9}{\`U}
\DeclareUnicodeCharacter{00DA}{\'U}
\DeclareUnicodeCharacter{00DB}{\^U}
\DeclareUnicodeCharacter{00DC}{\"U}
\DeclareUnicodeCharacter{00DD}{\'Y}
\DeclareUnicodeCharacter{00DF}{\ss}

\DeclareUnicodeCharacter{00E0}{\`a}
\DeclareUnicodeCharacter{00E1}{\'a}
\DeclareUnicodeCharacter{00E2}{\^a}
\DeclareUnicodeCharacter{00E3}{\~a}
\DeclareUnicodeCharacter{00E4}{\"a}
\DeclareUnicodeCharacter{00E5}{\aa}
\DeclareUnicodeCharacter{00E6}{\ae}
\DeclareUnicodeCharacter{00E7}{\c{c}}
\DeclareUnicodeCharacter{00E8}{\`e}
\DeclareUnicodeCharacter{00E9}{\'e}
\DeclareUnicodeCharacter{00EA}{\^e}
\DeclareUnicodeCharacter{00EB}{\"e}
\DeclareUnicodeCharacter{00EC}{\`i}
\DeclareUnicodeCharacter{00ED}{\'i}
\DeclareUnicodeCharacter{00EE}{\^i}
\DeclareUnicodeCharacter{00EF}{\"i}
\DeclareUnicodeCharacter{00F1}{\~n}
\DeclareUnicodeCharacter{00F2}{\`o}
\DeclareUnicodeCharacter{00F3}{\'o}
\DeclareUnicodeCharacter{00F4}{\^o}
\DeclareUnicodeCharacter{00F5}{\~o}
\DeclareUnicodeCharacter{00F6}{\"o}
\DeclareUnicodeCharacter{00F8}{\o}
\DeclareUnicodeCharacter{00F9}{\`u}
\DeclareUnicodeCharacter{00FA}{\'u}
\DeclareUnicodeCharacter{00FB}{\^u}
\DeclareUnicodeCharacter{00FC}{\"u}
\DeclareUnicodeCharacter{00FD}{\'y}
\DeclareUnicodeCharacter{00FF}{\"y}

\DeclareUnicodeCharacter{2013}{--}       
\DeclareUnicodeCharacter{2014}{---}      
\DeclareUnicodeCharacter{2018}{`}        
\DeclareUnicodeCharacter{2019}{'}        
\DeclareUnicodeCharacter{201C}{``}       
\DeclareUnicodeCharacter{201D}{''}       
\DeclareUnicodeCharacter{2022}{\textbullet}
\DeclareUnicodeCharacter{2026}{\ldots}   

\DeclareUnicodeCharacter{20AC}{\euro}    
\DeclareUnicodeCharacter{2122}{\texttrademark}

\begin{document}

\title{AgentSight: System-Level Observability for AI Agents Using eBPF}

\author{Yusheng Zheng}
\affiliation{%
  \institution{UC Santa Cruz}
  \city{Santa Cruz}
  \state{CA}
  \country{USA}}
\email{yzhen165@ucsc.edu}

\author{Yanpeng Hu}
\affiliation{%
  \institution{ShanghaiTech University}
  \city{Shanghai}
  \country{China}}
\email{huyp@shanghaitech.edu.cn}

\author{Tong Yu}
\affiliation{%
  \institution{eunomia-bpf Community}
  \country{China}
}
\email{yt.xyxx@gmail.com}

\author{Andi Quinn}
\affiliation{%
  \institution{UC Santa Cruz}
  \city{Santa Cruz}
  \state{CA}
  \country{USA}}
\email{aquinn1@ucsc.edu}

\sloppy
\begin{abstract}
    Modern software infrastructure increasingly relies on LLM agents for development and maintenance, such as Claude Code and Gemini-cli. However, these AI agents differ fundamentally from traditional deterministic software, posing a significant challenge to conventional monitoring and debugging. This creates a critical semantic gap: existing tools observe either an agent's high-level intent (via LLM prompts) or its low-level actions (e.g., system calls), but cannot correlate these two views. This blindness makes it difficult to distinguish between benign operations, malicious attacks, and costly failures. We introduce AgentSight, an AgentOps observability framework that bridges this semantic gap using a hybrid approach. Our approach, \emph{boundary tracing}, monitors agents from outside their application code at stable system interfaces using eBPF. AgentSight intercepts TLS-encrypted LLM traffic to extract semantic intent, monitors kernel events to observe system-wide effects, and causally correlates these two streams across process boundaries using a real-time engine and secondary LLM analysis. This instrumentation-free technique is framework-agnostic, resilient to rapid API changes, and incurs less than 3\% performance overhead. Our evaluation shows AgentSight detects prompt injection attacks, identifies resource-wasting reasoning loops, and reveals hidden coordination bottlenecks in multi-agent systems. AgentSight is released as an open-source project at \url{https://github.com/eunomia-bpf/agentsight}.
\end{abstract}


\settopmatter{printfolios=true}

\maketitle
\pagestyle{plain}

\section{Introduction}

The role of machine learning in systems is undergoing a fundamental shift from optimizing well-defined tasks, such as database query planning, to a new paradigm of \emph{agentic computing}. From a systems perspective, an AI agent couples a Large Language Model's (LLM) reasoning with direct access to system tools, granting it agency to perform operations like spawning processes, modifying the filesystem, and executing commands. This technology is being rapidly integrated into production environments, powering autonomous developer tools like Claude Code\cite{claudecode}, Cursor Agent\cite{cursor} and Gemini-CLI\cite{geminicli}, which can independently handle complex software engineering and system maintenance tasks. In essence, we are deploying non-deterministic ML systems, creating an unprecedented class of challenges for system reliability, security, and verification.

This paradigm shift creates a critical semantic gap: the chasm between an agent's high-level \emph{intent} and its low-level \emph{system actions}. Unlike traditional programs with predictable execution paths, agents use LLMs and autonomous tools to dynamically generate code and spawn arbitrary subprocesses. This makes it hard for existing observability tools to distinguish benign operations from catastrophic failures. Consider an agent tasked with code refactoring that, due to a malicious prompt it reads from external url in the search result when search for API documents, instead injects a backdoor (indirect prompt injection)\cite{indirect-prompt-inject}. An application-level monitor might see a successful "execute script" tool call, while a system monitor sees a \texttt{bash} process writing to a file. Neither can bridge the gap to understand that a benign intention has been twisted into a malicious action, rendering them effectively blind.

Current approaches are trapped on one side of this semantic gap. \emph{Application-level instrumentation}, found in frameworks like LangChain~\cite{langchain} and AutoGen~\cite{autogen}, captures an agent's reasoning and tool selection. While these tools see the \emph{intent}, they are brittle, require constant API updates, and are easily bypassed: a single shell command escapes their view, breaking the chain of visibility under a flawed trust model. Conversely, \emph{generic system-level monitoring} sees the \emph{actions}, tracking every system call and file access. However, it lacks all semantic context. To such a tool, an agent writing a data analysis script is indistinguishable from a compromised agent writing a malicious payload. Without understanding the preceding LLM instructions, the \emph{why} behind the \emph{what}, its stream of low-level events is meaningless noise.

We propose {boundary tracing} as a novel observability method designed specifically to bridge this semantic gap. Our key insight is that while agent internals and frameworks are volatile, the interfaces through which they interact with the world (the kernel for system operations and the network for communication) are stable and unavoidable. By monitoring from outside the application at these boundaries, we can capture an agent's high-level intent and its low-level system effects. We present \textbf{AgentSight}, a system that realizes boundary tracing using eBPF to intercept TLS-encrypted LLM traffic for intent and monitor kernel events for effects. Its core is a novel, two-stage correlation process: a real-time engine links an LLM response to the system bahavior it triggers, and a secondary "observer" LLM performs a deep semantic analysis on the resulting trace to infer risks and explain \emph{why} a sequence of events is suspicious. This instrumentation-free, framework-agnostic technique incurs less than 3\% overhead and effectively detects prompt injection attacks, resource-wasting reasoning loops, and multi-agent system bottlenecks.

In summary, our contributions are:

\begin{enumerate}
\item We introduce boundary tracing as a principled approach to AI agent observability that bridges the semantic gap by monitoring at stable system interfaces.
\item We present a novel engine that combines real-time, eBPF-based signal matching with LLM-based semantic analysis to provide deep, contextual understanding of agent behavior.
\item We demonstrate AgentSight's effectiveness in detecting prompt injection attacks, reasoning loops, and multi-agent coordination failures with sub-3\% overhead.
\end{enumerate}



\section{Background and Related Work}

This section outlines LLM agent architecture, reviews existing observability work to highlight the semantic gap, and introduces eBPF as our foundational technology.


\subsection{LLM Agent Architecture}
The agentic systems described in the introduction are typically implemented using a common architecture. These systems consist of three core components: (1) an LLM backend for reasoning, (2) a tool execution framework for system interactions, and (3) a control loop that orchestrates prompts, tool calls, and state management. Popular frameworks such as LangChain~\cite{langchain}, AutoGen~\cite{autogen}, Cursor Agent\cite{cursor}, Gemini-CLI\cite{geminicli} and Claude Code\cite{claudecode} all implement variations of this model. This architecture is what enables agents to dynamically construct and execute complex plans (e.g., autonomously writing and running a script to analyze a dataset) based on high-level natural language objectives. Multi-agent LLM systems have also emerged for software development, collaborative task-solving, simulating social behaviors in virtual worlds, and other tasks~\cite{guo2024survey,tran2025survey}.

\subsection{Observability for LLM Agent}

Existing approaches are siloed on one side of the semantic gap. Intent-side observability, supported by industry tools like Langfuse, LangSmith, and Datadog~\cite{Maierhofer2025Langfuse, langfuse, langsmith, Datadog2023Agents, helicone} and is unifying by standards from the OpenTelemetry GenAI working group~\cite{Liu2025OTel,Bandurchin2025Uptrace} and acadamics conceptual taxonomies~\cite{Dong2024AgentOps, Moshkovich2025Pipeline} under the AgentOps concept, excels at tracing application-level events but is fundamentally blind to out-of-process system \emph{actions}. Conversely, action-side observability with tools like Falco and Tracee~\cite{falco, tracee} offers comprehensive visibility into system calls but lacks the semantic context to understand an agent's \emph{intent}, failing to distinguish a benign task from a malicious one. A parallel line of research into reasoning-level and interpretability aims to make the agent's internal thought processes more transparent by reconstructing cognitive traces~\cite{Rombaut2025Watson} or enabling explanatory dialogues~\cite{Kim2025AgenticInterp}, but these work mainly focus on the llm itself, does not bridge the gap between the agent's internal reasoning and its external, low-level effects on the system.

\subsection{extended Berkeley Packet Filter (eBPF)}

To bridge the semantic gap, our approach requires a technology that can safely and efficiently observe both network communication and kernel activity. eBPF (extended Berkeley Packet Filter) is a fundamental advancement in kernel programmability that provides precisely this capability~\cite{brendangregg}. Originally designed for packet filtering, eBPF has evolved into a general-purpose, in-kernel virtual machine that powers modern observability and security tools~\cite{ebpfio,cilium}, and not limited to Linux\cite{zheng2025extending, windows-ebpf}. For AI agent observability, eBPF is uniquely suited because it allows observation at the exact boundaries where agents interact with the world, enabling both TLS interception for semantic \emph{intent} and syscall monitoring for system \emph{actions} with minimal overhead. Critically, its kernel-enforced safety guarantees, including verified termination and memory safety, make it ideal for production environments and provide a stable foundation for our solution~\cite{kerneldoc}.

\section{Design}

The design of AgentSight is guided by a single imperative: to bridge the semantic gap between an agent's intent and its actions. We achieve this through a novel observability method, boundary tracing, realized by a multi-signal correlation engine.

\subsection{Challenges}

The emergent and non-deterministic nature of AI agents fundamentally breaks traditional program observability paradigms, introducing two core challenges that original observability cannot address.

\textbf{Bridging the Semantic Gap Between Intent and Action}
The first and most significant challenge is bridging the vast semantic gap between an agent's high-level \textit{intent} and its low-level system \textit{actions}. Unlike conventional software, where intent is encoded in predictable source code, an agent's intent is expressed in natural language and interpreted by an LLM, creating dynamic "source code" that is generated at runtime. Consequently, it is impossible for a static analyzer to determine what an agent will do. For example, the intent "find and fix the bug in the authentication module" is semantically rich but operationally ambiguous, potentially resulting in a complex sequence of actions like reading files (\texttt{openat2}), compiling code (\texttt{execve} -> \texttt{gcc}), and running tests (\texttt{execve} -> \texttt{python}). This creates a critical observability problem: how can a monitoring system verify that the cascade of system calls is a legitimate fulfillment of the natural language intent? To solve this, an observer must move beyond simple pattern matching to gain a semantic understanding of the agent's goal, necessitating a new, llm based approach to interpret the correlated traces.

\textbf{Isolating the Causal Signal from High-Volume System Noise}
The second challenge stems from the agent's autonomy to use any tool necessary to achieve its goal, leading to an unpredictable and high-volume stream of system events. An agent might spawn shells, download scripts, or invoke compilers-processes that are not known ahead of time. This makes it exceedingly difficult to distinguish the agent's specific activity (the "signal") from the background noise of the operating system. Static, pre-configured filters, for instance, a rule to only monitor \texttt{git} commands, are inherently brittle and will fail the moment the agent uses \texttt{curl} and \texttt{bash} to achieve a similar outcome. Our design addresses this with an aggressive, dynamic in-kernel eBPF filter. By tracking process creation events (\texttt{fork}, \texttt{execve}), the filter builds a complete lineage tree of the agent's activity and dynamically applies rules in the kernel to only pass events from the agent or its descendants to userspace. This approach ensures that the entire causal chain is captured efficiently at the source, dramatically reducing overhead and providing a clean, high-fidelity signal for correlation and analysis.

\subsection{Boundary Tracing: A Principled Approach}

\begin{figure}[h!]
    \centering
    \includegraphics[width=\columnwidth]{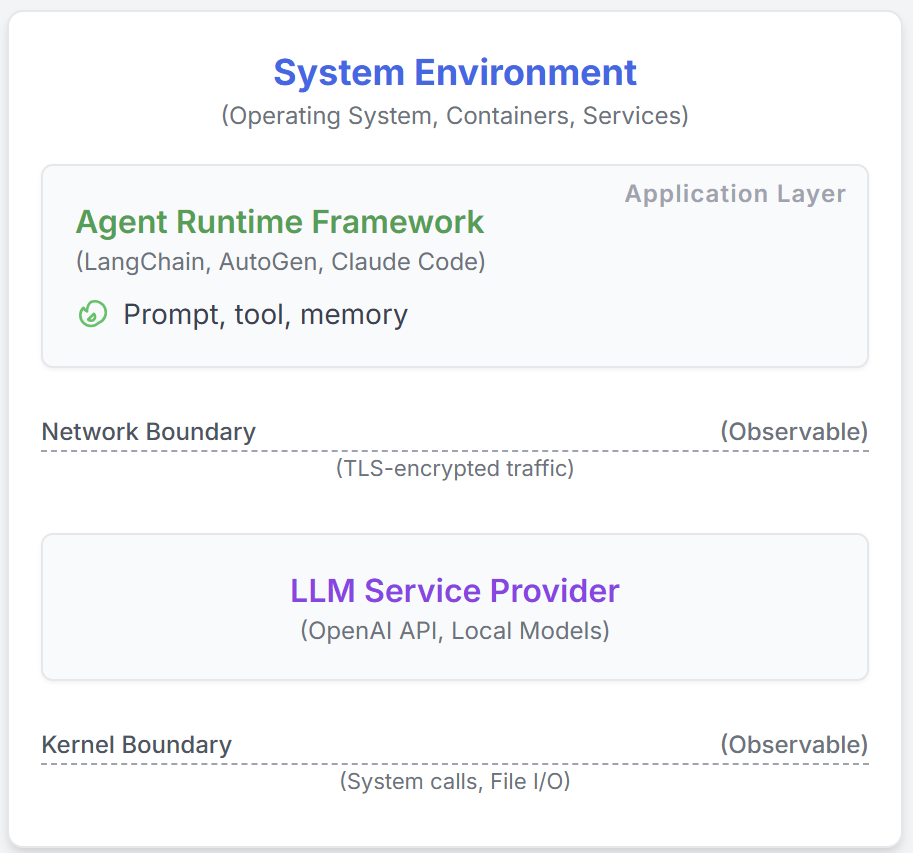}
    \caption{Agent Framework Overview}
    \label{fig:agent}
\end{figure}

Our key insight is that all agent interactions must traverse well-defined and stable system boundaries: the kernel for system operations and the network for external communications with LLM serving backends (Figure \ref{fig:agent}). By monitoring at these boundaries rather than within volatile agent code, we achieve comprehensive monitoring independent of implementation details. This approach enables Semantic Correlation, the ability to causally link high-level intentions with low-level system events. This is supported by two principles. First is Comprehensiveness, as kernel-level monitoring ensures no system action from process creation to file I/O goes unobserved, even across spawned subprocesses. Second is Stability, since system call ABIs and network protocols evolve far more slowly than agent frameworks, providing a durable, future-proof solution. This paradigm shifts the trust model from assuming a cooperative agent to enforcing observation at tamper-proof boundaries.

\subsection{System Architecture: Observing the Boundaries}

\begin{figure}[h!]
    \centering
    \includegraphics[width=\columnwidth]{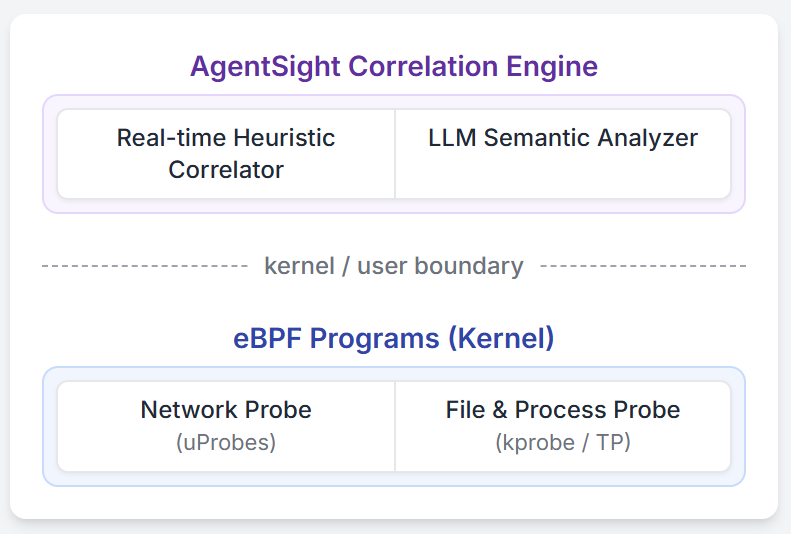} 
    \caption{\textbf{AgentSight System Architecture.}}
    \label{fig:architecture}
\end{figure}

AgentSight's architecture simultaneously taps into the two critical boundaries. As shown in Figure \ref{fig:architecture}, we use eBPF to place non-intrusive probes that capture a decrypted Intent Stream (LLM prompts/responses) from userspace SSL functions and an Action Stream (syscalls, process events) from the kernel. A userspace correlation engine then processes and joins these streams into a unified, causally-linked trace.

Several key components enable AgentSight to effectively bridge the semantic gap:

\textbf{eBPF for Safe, Unified Probing:} We chose eBPF for its production safety, high performance, and unified ability to access both userspace and kernel data streams. Our design intercepts decrypted data from the agent's interaction with LLM serving backend, which is more efficient and manageable than network-level packet capture or proxy-based solutions.

\textbf{Multi-Signal Causal Correlation Engine:} The core of our design is a correlation strategy that establishes causality between intent and action. We designed a multi-signal engine that relies on three key mechanisms: Process Lineage, which builds a complete process tree by tracking \texttt{fork} and \texttt{execve} events to link actions in child processes back to the parent agent; Temporal Proximity, which associates actions that occur within a narrow time window immediately following an LLM response; and Argument Matching, which directly matches content from LLM responses, such as filenames, URLs, or commands, with the arguments of subsequent system calls. Together, these signals enable AgentSight to definitively establish causal relationships between high-level intentions and low-level system operations across process boundaries.

\textbf{LLM-Powered Semantic Analysis:} To move beyond brittle, rule-based detection, we designed the system to use a secondary LLM as a reasoning engine. By prompting a powerful model with the correlated event trace, we leverage its ability to understand semantic nuance, infer causality in complex scenarios, and summarize findings in natural language. This "AI to watch AI" approach allows AgentSight to detect threats that do not match predefined patterns.

\section{Implementation}

AgentSight is implemented as a userspace daemon (6000 lines of Rust/C) orchestrating eBPF programs, with a TypeScript frontend (3000 lines) for analysis. It is designed for high performance, processing raw kernel event streams into correlated, human-readable data.

\subsection{Data Collection at the Boundaries}

Our eBPF probes capture the raw intent and action streams from the system. To capture semantic intent, an eBPF program with uprobes attaches to SSL\_read/SSL\_write in crypto libraries like OpenSSL to intercept decrypted LLM communications. Our userspace daemon implements a stateful reassembly mechanism to handle streaming protocols such as Server-Sent Events (SSE). To capture system actions, a second eBPF program uses stable tracepoints like sched\_process\_exec to build a process tree and kprobes to dynamically monitor relevant syscalls such as openat2, connect, and execve. To manage the high volume of kernel events without data loss, aggressive in-kernel filtering is applied to ensure only events from targeted agent processes are sent to userspace, minimizing overhead.

\subsection{The Hybrid Correlation Engine}

The Rust-based userspace daemon houses our two-stage correlation engine. The first stage consumes events from eBPF ring buffers and performs real-time heuristic linking. This streaming pipeline enriches raw events with context like mapping a file descriptor to a full path, maintains a stateful process tree, and applies the causal linking logic described in our design, using a 100-500ms window for temporal correlation. Once a coherent trace is constructed, the second stage formats it into a structured log for semantic analysis. This log is used to construct a detailed prompt for a secondary LLM, instructing it to act as a security analyst. The LLM's natural language analysis and confidence score become the final output of our system. A key challenge at this stage is managing the latency and cost of LLM analysis, which our system mitigates through asynchronous processing and robust prompt engineering.

\section{Evaluation}

Our evaluation is guided by two research questions: First, what is the performance overhead of AgentSight in realistic workflows? Second, how effectively does it bridge the semantic gap to detect critical security threats and performance pathologies, while also revealing complex dynamics in multi-agent systems?

\subsection{Performance Evaluation}

\begin{table}[h]
\centering
\caption{Overhead Introduced by AgentSight}
\label{tab:build-overhead}
\begin{tabular}{lrrr}
\toprule
Task & Baseline (s) & AgentSight (s) & Overhead \\
\midrule
Understand Repo & 127.98 & 132.33 & 3.4\% \\
Code Writing & 22.54 & 23.64 & 4.9\% \\
Repo Compilation & 92.40 & 92.72 & 0.4\% \\
\bottomrule
\end{tabular}
\end{table}

 We evaluated AgentSight on a server (Ubuntu 22.04, Linux 6.14.0) using Claude Code 1.0.62\cite{claudecode} with claude 4 as the test agent. The benchmarks focused on three real-world developer workflows using a tutorial repo\cite{ebpftutorial}: repository understanding with the \texttt{/init} command, code generation for bpftrace scripts, and full repository compilation with parallel builds. Each experiment was run 3 times with and without AgentSight to measure runtime overhead.
Table~\ref{tab:build-overhead} quantifies the runtime overhead of AgentSight across three developer workflows, with a average 2.9\% overhead.  

\subsection{Case Studies}

We evaluated AgentSight's effectiveness through case studies that demonstrate its ability to detect security threats, identify performance issues, and provide insights into complex multi-agent systems.

\subsubsection{Case Study 1: Detecting Prompt Injection Attacks}

We tested AgentSight's ability to detect indirect prompt injection attacks\cite{indirect-prompt-inject}. In our test, a software development agent is instructed to clone and build a C project. The project's README file directed the agent to a URL serving HTML with a hidden prompt. This prompt made the agent read and send /etc/passwd to a collection server, using a command cleverly disguised as a necessary step in the build process. AgentSight captured the full correlated attack chain: from the initial URL fetch to the final network exfiltration, with 521 events and merged into 37 events by  Correlation Engine. The observer LLM analyzed this trace, returned a high-confidence attack score, and concluded that the agent's actions were logically inconsistent with its stated goal. This result demonstrates how causally linking intent to system-level actions provides effective, context-aware threat detection.

\subsubsection{Case Study 2: Reasoning Loop Detection}

An agent attempting a complex task may enter an infinite loop due to a common tool usage error\cite{zhang2024breakingagents}
. We implement a research agent using crewai\cite{crewai} with gpt-4o-mini\cite{gpt4omini}, it repeatedly called a web search tool with incorrect arguments, received an error, but then failed to correct its mistake, retrying the exact same failing command. AgentSight's real-time monitors detect this anomalous resource consumption from a trace of API calls and passed it to the observer LLM. The LLM identified the root cause as a persistent tool error, noting the agent was caught in a "try-fail-re-reason" loop; it executed the same failing command, passed the identical error back to the reasoning LLM, and failed to learn from the tool's output. 

\subsubsection{Case Study 3: Multi-Agent Coordination Monitoring}

AgentSight monitored a team of 6 collaborating software development agents for our Github repo, using claude-code subagents\cite{subagents}, captured 3153 total events after Correlation Engine. For instance, frontend agent and test agent sometimes are blocked by sequential dependencies and numerous retry cycles caused by file locking contention during parallel development and testing tasks. The analysis demonstrated that while the agents developed some emergent coordination, separating the roles more clear could reduce total runtime and token cost. This reveals how boundary tracing uniquely captures multi-agent system dynamics that application-level monitoring cannot observe across process boundaries.

\section{Conclusion}

This paper introduced AgentSight to bridge the critical semantic gap between an AI agent's intent and its system-level actions using novel \textit{boundary tracing} approach. By leveraging eBPF, the system monitors network and kernel events without instrumentation, causally linking LLM communications to their system-wide effects via a hybrid correlation engine. Our evaluation shows AgentSight effectively detects prompt injection attacks, reasoning loops, and multi-agent bottlenecks with under 3\% performance overhead. This "AI to watch AI" provides a foundational methodology for the secure and reliable deployment of increasingly autonomous AI systems.


\bibliographystyle{ACM-Reference-Format}
\bibliography{ai}

\end{document}